\documentclass[twocolumn,prc,floatfix,preprintnumbers,%
superscriptaddress,nofootinbib]{revtex4}
\usepackage[utf8x]{inputenc}
\usepackage{mathrsfs}
\usepackage{amssymb}
\usepackage{amsmath}
\usepackage{graphicx}

\usepackage{float}

\usepackage{xcolor}
\definecolor{lcolor}{rgb}{0.5,0,0}
\definecolor{citcolor}{rgb}{0,0.3,0.0}

\usepackage[breaklinks,colorlinks,urlcolor=blue,citecolor=citcolor,linkcolor=lcolor]{hyperref}
\usepackage{mciteplus}

\newcommand{\der}{\mathrm{d}}
\newcommand{\rt}{{\mathbf{r}_T}}

\newcommand{\bt}{{\mathbf{b}_T}}

\newcommand{\pt}{{\mathbf{p}_T}}

\newcommand{\gev}{\ \textrm{GeV}}
\newcommand{\tev}{\ \textrm{TeV}}

\newcommand{\qso}{Q_{\mathrm{s}0}}
\newcommand{\lqcd}{\Lambda_{\mathrm{QCD}}}
\newcommand{\as}{\alpha_{\mathrm{s}}}

\begin{document}

\author{Heikki Mäntysaari}
\author{Hannu Paukkunen}
\affiliation{
Department of Physics,  University of Jyväskylä, %
 P.O. Box 35, 40014 University of Jyväskylä, Finland
}
\affiliation{
Helsinki Institute of Physics, P.O. Box 64, 00014 University of Helsinki, Finland
}

\title{
Saturation and forward jets in proton-lead collisions at the LHC
}


\preprint{}

\begin{abstract}
We investigate the forward-jet energy spectrum within the Color Glass Condensate framework at 5\,TeV center-of-mass energy. In particular, we focus on the kinematic range covered by the CMS-CASTOR calorimeter. We show that our saturation-model calculations are compatible with the CASTOR measurements and that to optimally reproduce the data, effects of multi-parton interactions need to be included. We predict a significant nuclear suppression -- reaching down to 50\% at the lowest considered jet energies $E_{\rm jet} \sim 500 \, {\rm GeV}$. 
\end{abstract}

\maketitle

\section{Introduction}

The non-linear phenomena of Quantum Chromodynamics (QCD) are expected to have a significant effect on high-energy interactions of protons and heavier nuclei \cite{Iancu:2003xm}.  In particular, at sufficiently small momentum fraction $x$ one expects to enter the saturation region, where the  parton densities reach their maximally allowed value and the effect known as \emph{parton saturation} kicks in. The formation mechanism and properties of this dynamically generated confined many-body system of partons are not known in detail, partly due to an insufficient amount of available experimental data probing truly small-$x$ structure of hadrons. Not only is the understanding of this high-density regime an intriguing task on its own, but also a necessary input for an accurate interpretation of the measurements performed in heavy-ion collisions.

A convenient theoretical framework to describe QCD in the high-energy limit, where saturation effects can become significant, is provided by the Color Glass Condensate (CGC) effective field theory~\cite{Gelis:2010nm,Albacete:2014fwa}. The perturbative CGC evolution equations, like the  Balitsky-Kovchegov (BK) equation~\cite{Balitsky:1995ub,Kovchegov:1999yj} describe the energy evolution of the hadron structure, and predict the $x$ dependence of cross sections.

Today, the most precise picture of the partonic structure of protons is provided by deep inelastic scattering measurements performed at HERA~\cite{Aaron:2009aa,Abramowicz:2015mha} that probe the proton structure down to $x\sim 10^{-5}$. These precise measurements are known to be in a good agreement with saturation model calculations~\cite{Albacete:2010sy,Rezaeian:2012ji,Mantysaari:2018nng} performed at leading logarithmic accuracy, resumming contributions of $\as \ln 1/x$. However, the same dataset is equally well described by collinearly factorized parton distribution functions (PDFs), whose $x$ dependence is parametrized at low $Q^2$ and obtained at higher $Q^2$ by solving the perturbative DGLAP evolution equations~\cite{Gribov:1972ri,Gribov:1972rt,Altarelli:1977zs,Dokshitzer:1977sg}. Recently, it has also been argued that a resummation of $\ln 1/x$ terms improves the description of HERA data in PDF fits~\cite{Ball:2017otu}, though similar effects can be achieved by an appropriate choice for the factorization and renormalization scales \cite{Hou:2019qau}. Nevertheless, explicit parton saturation which would be a result of non-linear QCD dynamics is absent in these analyses and thus it seems that the HERA data alone cannot confirm nor disprove the presence of saturation phenomena at perturbative interaction scales.

As the gluon densities are enhanced by roughly $A^{1/3}$, replacing protons by heavier ions should render the non-linear QCD effects more easily observable. Future electron-ion colliders (EICs) in the US~\cite{Accardi:2012qut} and at CERN~\cite{AbelleiraFernandez:2012cc} are designed to probe the small-$x$ structure of heavy nuclei with electron beams very precisely \cite{Aschenauer:2017oxs,HannuPaukkunenfortheLHeCstudygroup:2017ric}. Before such machines are realized, large gluon densities in nuclei can be probed studying facilities production at forward rapidities in proton-nucleus collisions, as the Bjorken-$x$ in the target nucleus scales like $x_A \sim e^{-y}/\sqrt{s_{NN}}$. Particle production at forward rapidities with $\sqrt{s_{NN}}=200 \gev$ center-of-mass (c.m.) energy has been measured at RHIC in deuteron-gold collisions~\cite{ Adams:2006uz,Adler:2004eh,Arsene:2004ux,Tribedy:2010ab}, and the results are compatible with CGC calculations at leading order~\cite{Albacete:2010bs,Lappi:2013zma}. As the c.m. energies at the LHC are much higher, $\sqrt{s_{NN}}=5020,8160 \gev$ in proton-lead collisions, a similar kinematic configuration as what is reached at RHIC with $y \gg 0$ is probed already at midrapidity, $y \sim 0$. Convincingly, e.g. the ALICE measurements for pion production~\cite{Acharya:2018hzf} are indeed compatible with the saturation-model predictions. 

Particle production at the LHC in the forward direction will probe much smaller values of $x$ in the nucleus than what is achievable at RHIC. Currently, the most decisive small-$x$ data are arguably the D- and B-meson measurements by the LHCb collaboration \cite{Aaij:2017gcy,Aaij:2019lkm} which both show that the p-Pb cross sections at $y \gg 0$ are clearly suppressed with respect to the p-p baseline. Out of these two the D-meson production should be the most sensitive to the non-linear dynamics due to the low interaction scale involved. The measurements are indeed in fair agreement with the CGC predictions~\cite{Ducloue:2016ywt}. However, as accurate description can be obtained~\cite{Eskola:2019bgf} in the collinearly factorized framework when an appropriate general-mass scheme~\cite{Helenius:2018uul} and nuclear PDFs~\cite{Paukkunen:2017bbm,Paukkunen:2018kmm} are used. So it looks like a tie for collinear factorization vs. saturation calculations also in the case of heavier ions. 

Out of other observables, 
one can look at quarkonium production which has been measured at forward rapidities at the LHC (see e.g.~\cite{Aaij:2011jh,Abelev:2014qha,Chatrchyan:2011kc,Abelev:2013yxa,Aaij:2013zxa,Abelev:2014oea}), and analyzed in the saturation picture extensively e.g. in Refs.~\cite{Fujii:2013gxa,Ducloue:2015gfa,Ducloue:2016pqr,Ma:2014mri,Ma:2017rsu}. The disadvantage of these processes is that the results are more difficult to interpret due to the complexity of the $J/\Psi$ or $\Upsilon$ meson formation that has to be modelled in the calculations. Some extremely low-$p_T$ data are also available, see Refs.~\cite{Adriani:2012ap} and \cite{Goncalves:2012bn}. In future, e.g. direct photon measurements \cite{Peitzmann:2014isa,Boettcher:2019kxa} could lead to further insight \cite{Helenius:2014qla}.

Recently, it has also become possible to measure photon-nucelus scattering in ultra peripheral heavy ion collisions (UPCs) at RHIC and at the LHC, where the impact parameter is so large that strong interactions are suppressed~\cite{Bertulani:2005ru,Klein:2019qfb}. So far exclusive vector-meson production has been the focus in UPC studies at RHIC and at the LHC (see e.g.~\cite{Afanasiev:2009hy,Abbas:2013oua,Khachatryan:2016qhq,LHCb:2018ofh}).  The advantage of UPCs is that it becomes possible to study nuclear DIS at high energies (albeit only in the photoproduction region) before an EIC is realized. Clear signals of significant nuclear suppression have been seen in these measurements~\cite{Lappi:2013am,Guzey:2013qza}.

A potentially useful place to look for further answers is particle production at very forward rapidities covered by the CASTOR calorimeter of the CMS experiment. This apparatus spans the pseudorapidity region $5.2 < \eta < 6.6$ in the laboratory frame which, in the proton-lead run setup, is boosted by $y_\text{shift}=0.465$ to the direction of the proton beam. 
Interestingly, the first jet-energy spectra in proton-lead collisions measured using CASTOR is recently published~\cite{Sirunyan:2018ffo}.
Thanks to the very forward rapidity and large c.m. energy, Bjorken-$x$ values down to $x_A\sim 10^{-6}$ are probed. A disadvantage is that as no tracking is in place, only the total-energy production can be measured. In addition, there is no segmentation in $\eta$ and consequently it is not possible to study the rapidity dependence of jet production. This also means that the jets can be fairly fat and thus contributions of multi-parton interactions are potentially significant. Using inclusive jets as a probe for small-$x$ structure of heavy ions is in contrast to the usual paradigm of using jets to probe the large-$x$ parton content of nuclei \cite{Eskola:2013aya,Eskola:2019dui}. 

In this article, we present calculations for the jet-energy spectra within the CASTOR kinematics. The predictions are based on the CGC framework that was applied in Ref.~\cite{Lappi:2013zma} and which successfully describes inclusive particle production at RHIC and at the LHC. We show that the jet-energy spectra computed in the CGC framework are compatible with the ones measured by CASTOR. Additionally, we argue that there is a significant nuclear suppression present which we attribute to the saturation effects. 

The paper now continues as follows: In Sec.~\ref{sec:sinc}  we first review the calculation of single-parton production cross sections 
in the CGC picture. This will serve as our proxy to the jet cross section. In Sec.~\ref{sec:kinematics} we discuss the jet production in CASTOR kinematics before underscoring the role of multi-parton scattering in Sec.~\ref{sec:dps}. Numerical results are presented in Sec.~\ref{sec:results}.

\section{Forward jet production in CGC}
\label{sec:sinc}
In this work, we compute the parton-production cross section and assume that the produced parton creates a jet whose total energy equals the energy of the parton. The forward particle production in high-energy proton-nucleus collisions can be described using the so called hybrid formalism, in which the dilute moderate-$x$ probe is described in terms of collinear parton distribution functions, and the parton-target scattering is computed from the CGC taking into account multiple scattering with the target. 

\subsection{Proton-proton collisions}

In case of a proton target, the leading-order parton production cross section at rapidity $y$ in the center-of-mass frame reads~\cite{Dumitru:2002qt,Dumitru:2005gt}
\begin{equation}
\label{eq:hybrid}
\frac{\der \sigma^{p+p \to i+X}}{\der y \der^2 \pt} = \frac{\sigma_0}{2} \frac{1}{(2\pi)^2} x f_i(x,\mu^2) S(\pt,x),
\end{equation}
where $xf_i(x,\mu^2)$ is the (leading order) parton density function for a parton $i$ at scale $\mu^2$. In this work we use the CTEQ6 parton distribution function~\cite{Pumplin:2002vw} and take $\mu=\max\{|\pt|, 1.3\gev\}$. The proton transverse area $\sigma_0/2$ is measured in DIS experiments, and is different from the inelastic proton-proton cross section $\sigma_\text{inel}$ which connects invariant yield and cross section~\cite{Lappi:2013zma}. The parton-target scattering is described in terms of the dipole amplitude $N$ Fourier transformed to momentum space: 
\begin{equation}
S(\pt,x) = \int \der^2 \rt e^{i \pt \cdot \rt} (1-N(\rt,x)).
\end{equation}

The energy (Bjorken-$x$) dependence of the dipole amplitude $N$ is obtained by solving the BK equation with running coupling~\cite{Balitsky:2006wa}, and the initial condition for the evolution (dipole amplitude at $x=x_0=0.01$) is obtained by fitting the HERA structure-function data as in Ref.~\cite{Lappi:2013zma}. The initial condition is parametrized as 
\begin{equation}
N_{x_0}(\rt) = 1 - \exp\left[ -\frac{\rt^2 \qso^2}{4} \ln \left(  \frac{1}{|\rt| \lqcd} + e_c\cdot e \right) \right]
\end{equation}
with the parameters fit to HERA structure function data being $\qso^2=0.06\gev^2$, $e_c=18.9$ and $\sigma_0/2=16.36$ mb.  When the parton flavor $i$ is gluon, the dipole is evaluated in the adjoint representation as $N_A = 2N - N^2$. 

Beyond leading order, Eq.~\eqref{eq:hybrid} can potentially acquire numerically significant corrections. Recently, there has been rapid progress in developing the theory to next-to-leading order accuracy, including the NLO BK evolution equation with numerical solution~\cite{Balitsky:2008zza,Lappi:2016fmu,Ducloue:2019ezk}. The particle-production cross section has also been computed at NLO accuracy~\cite{Chirilli:2012jd}, but phenomenological applications at this order are still problematic~\cite{Ducloue:2017dit,Ducloue:2016shw,Watanabe:2015tja,Altinoluk:2014eka,Stasto:2013cha}. On the other hand, as discussed in the Introduction, the leading order calculations have been successful in describing particle spectra and nuclear suppression factors at RHIC and at the LHC. Consequently, we limit ourselves here to the leading-order accuracy, thereby resumming contributions $\sim \as \ln 1/x$ to all orders by solving the BK evolution equation. 

\subsection{Proton-nucleus collisions}

Let us then consider proton-nucleus collisions. 
The dipole amplitude for the dipole-nucleus scattering is obtained as follows. First, the initial condition for the BK evolution at a given impact parameter $\bt$ is obtained by generalizing the dipole-proton scattering amplitude by applying the Glauber model. Then, the impact parameter independent BK evolution is performed separately at every impact parameter.
Neglecting the impact-parameter dependence from the BK evolution makes it possible to avoid long-distance Coulomb tails that should be regulated by confinement-scale physics, see e.g. Refs.~\cite{Berger:2011ew,Mantysaari:2018zdd} for more details. Following Ref.~\cite{Lappi:2013zma}, the initial condition at a given impact parameter $\bt$ is given as
\begin{multline}
\label{eq:nuke-ic}
N^A_{x_0}(\rt,\bt) = 1 - \exp\left[ -A T_A(\bt) \frac{\sigma_0}{2} \frac{\rt^2 \qso^2}{4} \right. \\
	\left. \times \ln \left( \frac{1}{|\rt| \lqcd} + e_c \cdot e \right) \right].
\end{multline}
Note that there are no parameters besides the Woods-Saxon nuclear density $T_A(\bt)$ (normalized such that $\int \der^2 \bt T_A(\bt)=1$) related to the nuclear structure.

In proton-nucleus scattering the cross section is obtained by integrating the invariant yield over the impact parameter $\bt$. The invariant yield at given impact parameter reads~\cite{Lappi:2013zma}
\begin{equation}
\label{eq:hybrid_pa}
\frac{\der N^{p+A \to i+X}(\bt)}{\der y \der^2 \pt} =  \frac{1}{(2\pi)^2} x f_i(x,\mu^2) S(\pt,x,\bt).
\end{equation}
As our centrality classes from the Optical Glauber model are not the same as the experimental centrality classes, we only consider minimum-bias collisions here. When computing the 
integral over impact parameters,  we have to describe the dilute edge of the nucleus. In order to avoid unphysically fast growth of the nuclear area, we follow Ref.~\cite{Lappi:2013zma} and assume that the particle production yield in proton-nucleus collisions  is the proton-proton yield scaled by $N_\text{bin}(\bt) = A T_A(\bt) \sigma_\text{inel}$ in the region where the saturation scale of the nucleus would fall below that of the proton. This is a natural approximation as it results in a nuclear suppression factor $R_{pA}=1$ in this dilute region. In this work we take $\sigma_\text{inel}=70$ mb at $\sqrt{s_{NN}}=5.02$ TeV.

\section{Kinematics of jets at CASTOR}
\label{sec:kinematics}

The CASTOR calorimeter at CMS is used to measure the forward jet production cross section as a function of jet energy  in the laboratory frame. It covers the pseudorapidity region $5.2 < \eta_\text{lab} < 6.6$. We can compute the differential cross section as
\begin{multline}
\label{eq:jet_spectra}
	\frac{\der \sigma}{\der E} \approx   \frac{1}{\Delta E} \int_{p_{T,\text{min}}} \hspace{-1em} \der^2 \pt \der y \, \theta\left(E_\text{lab} \in E \pm \frac{\Delta E}{2} \right) \\
	\times \theta(5.2 < \eta_\text{lab} < 6.6) 
	  \frac{\der \sigma^{p+p(A) \to i + X}}{\der^2 \pt \der y},
\end{multline}
where the parton-level cross section  in proton-proton(nucleus) collision $\sigma^{p+p(A) \to i+X}$ is summed over quarks $i=u,d,s,c,b$, antiquarks and gluons. The minimum transverse momentum is chosen to be $p_{T,\text{min}}=1.0\gev$ and the sensitivity to this choice is studied in Appendix~\ref{app:ptcut}. The measurement functions force the jets inside the CASTOR acceptance and to be inside the given energy bin. In our numerical calculations we will use the bin width $\Delta E=10\gev$. The jet energy $E_\text{lab}$ (assumed to be equal to the parton energy) is measured in the laboratory frame which is boosted by  $y_\text{shift}=0.465$ compared to the center of mass frame. From the jet transverse momentum and rapidity, we also compute the pseudorapidity $\eta_\text{lab}$ in the laboratory frame.

When relating rapidity and pseudorapidity, and when computing the jet energy, it becomes necessary to define the mass of the jet $m$. In this work we assume that the particle production is dominated by lightest hadrons (pions), and we use an effective mass $m = 0.2\gev$. As we will show in the Appendix~\ref{appendix:mass}, our result are not sensitive to the exact numerical value of $m$. The jet energy in the laboratory frame reads
\begin{equation}
\label{eq:jete}
	E =  m_T \cosh (y_\text{lab}),
\end{equation}
where the transverse mass is $m_T=\sqrt{m^2+\pt^2}$. 
The Bjorken $x$ values probed in these collisions are
\begin{align}
	x_p &= \frac{m_T}{\sqrt{s_{NN}}} e^y \\
	x_A &= \frac{m_T}{\sqrt{s_{NN}}} e^{-y}.
\end{align}
Here the jet rapidity $y$ is measured in the center of mass frame. In proton-nucleus collisions, the probed values of $x_A$ range from $x_A\sim 10^{-6}$ to $x_A\sim 10^{-5}$ in the CASTOR acceptance at $500 \gev < E < 3000 \gev$. Similarly, the momentum fractions in the probe are $x_p \sim 0.1 \dots 0.5$.

\section{Multi-parton scattering}
\label{sec:dps}
\subsection{Proton-proton collisions}

CASTOR only measures the total energy, and it is possible that the two independently produced jets are merged into one. As there is no segmentation in rapidity in CASTOR, this happens if the two produced jets are close to each other in azimuthal angle. At next-to-leading order, one would include dijet production where the probing parton emits a gluon before or after the scattering~\cite{Marquet:2007vb}. This perturbative contribution is predominantly back-to-back (although saturation effects suppress the $\Delta \varphi=\pi$ contribution~\cite{Lappi:2012nh}). Given the fact that NLO calculations of single-jet production in the CGC framework are not yet developed to the level where straightforward phenomenological applications are possible, we limit ourselves to the leading order accuracy in this work.

Instead, we take into account multiple parton interactions that can produce multiple jets at forward rapidity, which are measured as one if they are close to each other in azimuth. We assume that the probability to have $k$ partonic scattering processes is given by the Poisson distribution
\begin{equation}
p_k(\bt) = e^{-T(\bt) \sigma} \frac{[T(\bt) \sigma]^k}{k!},
\end{equation}
where $T(\bt)$ is the transverse density profile of the target, and $\sigma$ is the integrated jet production cross section:
\begin{equation}
\sigma = \int \der E \frac{\der \sigma}{\der E}.
\end{equation}
The final measured cross section is obtained by summing over all possible number of parton-target scattering processes. In parton-proton scattering we assume that the impact parameter dependence factorizes, and the cross section  reads~\cite{Eskola:1988yh}
\begin{multline}
\label{eq:master}
\frac{\der \sigma_\text{MPI}^{pp}}{\der E} = \int \der^2 \bt \sum_{j=1}^\infty p_k(\bt) \prod_{i=1}^j \left[ \frac{1}{\sigma }\int \der E_i \frac{\der \sigma}{\der E_i} \right] \\ \times \sum_{\text{measured}} \delta(E - E_\text{measured}). 
\end{multline}
In Sec.~\ref{sec:pa_mpi} we will consider proton-nucleus collisions and generalize this result to the case where the impact parameter dependence does not factorize.
Here, $\sum_\text{measured}$ refers to the summation over all possible combinations of produced partons. Implicitly, this also requires that when one sums over parton energies $E_1, \dots, E_n$, then these partons must be close in azimuthal angle in order to be measured as one jet in the calorimeter. Additionally, the other $k = j - n$ partons must not be merged with them.

Let us first consider the case where there is no merging, and the measured energy is just the energy of the single parton. This can be any of the partons produced in the scattering process, but the other $k$ partons must be at least distance $R=0.5$ away (otherwise they would be merged, which is not allowed). Additionally, there are $k+1$ ways to choose which parton is our measured jet. Taking this into account, the cross section becomes
\begin{multline}
\label{eq:pp_1jet}
\frac{\der \sigma^{pp, 1}_\text{MPI}}{\der E} =  \int \der^2 \bt e^{-\sigma T(\bt)} T(\bt)   \\
\times \frac{\der \sigma}{\der E} 
 \sum_{k=0} \frac{[\sigma T(\bt)]^k}{k!} C_1(k)
\end{multline}
where $C_1(k) = \left(1-\frac{2R}{2\pi}\right)^{k}$ is the probability that, when producing $1+k$ jets, the $k$ additional ones are not merged with jet $1$. 
If we put $C_1(k)=1$ here, corresponding to the case where we accept all possible azimuthal angles for the produced jets, the inclusive cross section summed over any number of multi parton interactions becomes $\der \sigma/\der E$.

The case where two of the jets are merged into one is computed similarly.  From  Eq.~\eqref{eq:master} we can directly write
\begin{multline}
\label{eq:pp_2jet}
\frac{\der \sigma^{pp,2}_\text{MPI}}{\der E} = \frac{1}{2!} \int \der^2 \bt e^{-\sigma T(\bt)} T^2(\bt)   \\
\times  \sum_{k=0} \frac{[\sigma T(\bt)]^k}{k!} C_2(k) \\
\times  \int \der E_1 \der E_2 \delta(E_1+E_2 - E)  \frac{\der \sigma}{\der E_1} \frac{\der \sigma}{\der E_2}.
\end{multline}
 As before, the coefficients $C_2(k)$ describe the probability that when producing $2+k$ jets, the $2$ are produced within a single jet cone with radius $R$ and consequently measured as one, and that the other $k$ are far enough not to be merged with these. 
  The generalization to $n=3$ merged jets and beyond is straightforward. 
  The numerical values for coefficients $C_n(k)$ are calculated as follows. First, with probability $[2R/(2\pi)]^{n-1}$ we sample $n$ jets within a single jet cone (radius $R$). Then, the azimuthal angles for the other $k$ jets are sampled uniformly, and the event is accepted if none of the extra $k$ jets are closer than $R$ to any of the merged jets.\footnote{Our jet merging alogirthm is only based on the azimuthal angle of the independently produced jets, and does not result in exactly same jet merging as the full anti-$k_T$ argorithm used in the experimental analysis. However, as we do not want to generate full monte carlo event samples, we stick with the simple approach which we expect to be a reasonable approximation of the full jet finding process.} 
  The numerically obtained values at small $k$ are shown in Table.~\ref{table:cnk}.
 Note that in the case where there are no limits for the azimuthal distribution of jets that are not merged, we would have $C_n(k)=(2R/(2\pi))^{n-1}$, and ${\der \sigma^{pp,2}_\text{MPI}}/{\der E}$ would reduce to the standard double-parton scattering (DPS) cross section \cite{dEnterria:2017yhd}, 
\begin{align}
\label{eq:dps_pp}
\frac{\der \sigma^{pp,2}_\text{MPI}}{\der E}
\xrightarrow{C_2(k)=\frac{R}{\pi}}
&
\int \der E_1 \der E_2 \delta(E_1+E_2 - E) \\
& \times \left(\frac{2R}{2\pi} \right) \frac{1}{2\sigma_\text{eff}} \frac{\der \sigma}{\der E_1} \frac{\der \sigma}{\der E_2},
\nonumber
\end{align}
with $\sigma_\text{eff}^{-1} = \int \der^2 \bt T^2(\bt)$. Here the factor $2R/2\pi$ takes into account the fact that the DPS process is included only if the two independently produced jets are close to each other in azimuthal angle. We will refer to this approximation as \emph{naive}.

\begin{table}[tp]
\begin{center}
\begin{tabular}{c|ccccc}
& $k=0$ & $k=1$ & $k=2$ & $k=3$ & $k=4$ \\
\hline
$n=1$ & 1 & 0.841 & 0.707 & 0.594 & 0.500\\
$n=2$ & 0.159 & 0.125 & 0.099 & 0.078 & 0.062\\
$n=3$ & 0.025 & 0.019 & 0.015 & 0.011 & 0.009 \\
\end{tabular}
\end{center}
\caption{Coefficients $C_n(k)$ describing the probability that in the production of $n+k$ jets, $n$ are merged as one when the cone radius is $R=0.5$.}
\label{table:cnk}
\end{table}%

When parton production at very forward rapidities is considered, the Bjorken-$x$ of the probing proton can become large (of the order of $x\sim 0.5$ in the CASTOR kinematics), and one has to take into account the kinematical constraint that $x_1 + x_2 < 1$. We neglect the longitudinal momentum carried by the other $k$ produced jets that are not merged, as the cross section is dominated by jets carrying small energy fractions. 

In the case where we produce two jets that are merged into one, we implement the kinematical constraint $x_1+x_2 < 1$ by writing the two jet production cross section in terms of double parton distribution function $D_{ij}(x_1,x_2)$ following Ref.~\cite{Lappi:2012nh}\begin{multline}
\label{eq:dpdf}
D_{ij}(x_1,x_2) = \frac{1}{2} x_1 x_2 \left[ f_i(x_1) f_j\left(\frac{x_2}{1-x_1} \right) \right. \\
\left. + f_i\left( \frac{x_1}{1-x_2}  \right) f_j(x_2) \right],
\end{multline}
with the condition $f_i(x)=0$ for $x>1$ to guarantee the energy conservation (the effect of energy conservation is analyzed in Appendix~\ref{appendix:kc}). The scale at which the PDFs are evaluated is set by the average transverse momenta of partons, and $f_i(x)$ refers to $f_i(x, \mu_i^2)$ with $\mu_i=\max\{(|\pt_1|+|\pt_2|)/2, 1.3\gev\}$. 
Note that when $x_i,x_j\ll1$ this reduces to the factorized assumption $D_{ij}(x_1,x_2) = x_1 f_i(x_1) x_2 f_j(x_2)$. With this definition, the double-parton production cross section in proton-proton collisions becomes
\begin{multline}
\label{eq:dps_eint_dpdf}
\int \der E_1 \der E_1 \delta(E-E_1-E_2) \frac{\der \sigma}{ \der E_1} \frac{\der \sigma}{\der E_2} \\
\to \left( \frac{\sigma_0}{2} \right)^2 \int \der y_1 \der y_2 \der^2 \pt_1 \der^2 \pt_2 \frac{1}{(2\pi)^4} D_{ij}(x_1,x_2, \bt) \\
\times  \theta(5.2 < \eta_{1,\text{lab}}, \eta_{2,\text{lab}} < 6.6) \delta(E_1 + E_2 - E)  \\  
	\times   S(\pt_1,x_1, \bt) S(\pt_2,x_2, \bt)	,
\end{multline}
with summation over parton flavors $i,j$ having rapidities $y_i$ and momenta $\pt_i$ in the center of mass frame, and energies $E_i$ and pseudorapidities $\eta_{i,\text{lab}}$ are given in the laboratory frame (which is the same as the CMS frame in the proton-proton collisions).
Note that the leading-order single parton production cross section used in this work does not include a perturbatively generated double-parton scattering contribution, so no additional subtraction is needed to  avoid double counting with the above DPS cross section result~\cite{Diehl:2017kgu} (see also related discussion in Ref.~\cite{Lappi:2012nh}). Generalizing multi-parton distribution function Eq.~\eqref{eq:dpdf} and cross section \eqref{eq:dps_eint_dpdf} to the case of three (or more) merged jets is now straightforward.

In proton-proton collisions, the  effective cross section is experimentally measured to be $\sigma_\text{eff} \approx 15$ mb~\cite{Abazov:2009gc}. This allows us to identify the effective cross section to be the proton size: $\sigma_\text{eff} = \sigma_0/2$ (note that the fitted value for $\sigma_0/2$ is very close to the experimentally determined $\sigma_\text{eff}$). This is achived with a Gaussian density profile $T_p(\bt)=1/(2\pi B_p) e^{-b^2/(2B_p)}$ with $B_p=3.34 \gev^{-2}$, slightly less than $B_p \approx 4 \gev^{-2}$ suggested by the exclusive vector meson production data from HERA~\cite{Chekanov:2004mw,Aktas:2005xu}.

\subsection{Proton-nucleus collisions}
\label{sec:pa_mpi}
In proton-nucleus collisions, when a large-$x$ parton from the proton probes the structure of the large target nucleus, the parton-nucleus interaction depends on the impact parameter of the collision, and this dependence does not factorize, see Eq.~\eqref{eq:hybrid_pa}. This requires us to generalize the above discussion to the case where the parton-target interaction depends explicitly on the impact parameter.

In this case we generalize the MPI cross section \eqref{eq:master} result as follows:
\begin{multline}
\label{eq:master_pA}
\frac{\der \sigma_\text{MPI}^{pA,1}}{\der E} = \int \der^2 \bt \sum_{k=1}^\infty e^{-\sigma^{pA} T_A(\bt)} \frac{[\sigma^{pA} T_A]^k}{k!}  C_n(k)  \\
	\times  \prod_{i=1}^k\left[\frac{1}{N_\text{tot}(\bt)} \int \der E_i \frac{\der N^{pA}(\bt)}{\der E_i} \right] \\
	 \times \sum_\text{measured} \delta(E - E_\text{measured}).
\end{multline}
Here $N_\text{tot}(\bt) = \int \der E \frac{\der N^{pA}(\bt)}{\der E}$ and the parton production yield at given impact parameter in proton-nucleus collisions is given in Eq.~\eqref{eq:hybrid_pa}. Note that if the impact parameter dependence factorizes from the invariant yield, one recovers \eqref{eq:master}. The integrated cross section $\sigma^{pA}$ is obtained by integrating the invariant yield \eqref{eq:hybrid_pa} over impact parameter and CASTOR kinematics. One can easily check that if we impose no constraints on the azimuthal distribution of jets and take the dilute limit, the standard Optical-Glauber model result for single-jet production is obtained.

In the case where we do not allow other $k$ jets to be merged with the produced jet, the cross section now becomes
\begin{multline}
\frac{\der \sigma^{pA, 1}_\text{MPI}}{\der E} = \int \der^2 \bt e^{-\sigma^{pA} T_A(\bt)} \sigma^{pA} T_A(\bt) \\
\times  \sum_{k=0}^\infty  \frac{[\sigma^{pA} T_A]^k}{k!} C_1(k) 
 \left[\frac{1}{N_\text{tot}(\bt)} \frac{\der N^{pA}(\bt)}{\der E} \right]. 
\end{multline}
Similarly, in the case where two jets are merged, we can generalize Eq.~\eqref{eq:pp_2jet} and obtain
\begin{multline}
\label{eq:pa1_nonuke_1}
\frac{\der \sigma^{pA,2}_\text{MPI}}{\der E} = \frac{1}{2!} \int \der^2 \bt e^{-\sigma^{pA} T_A(\bt)} [\sigma^{pA} T_A(\bt)]^2   \\
\sum_{k=0} \frac{[\sigma^{pA} T_A]^k}{k!} C_2(k) \frac{1}{N_\text{tot}^2(\bt)} \\
\times \int \der E_1 \der E_2 \delta(E_1 + E_2 - E) \frac{\der N^{pA}(\bt)}{\der E_1} \frac{\der N^{pA}(\bt)}{\der E_2},
\end{multline}
where the kinematical constraint $x_1+x_2<1$ is added similarly as in case of proton-protons scattering. In the dilute limit one again recovers the standard double-parton scattering result \eqref{eq:dps_pp}. It is now easy to see how this generalizes to the case where any number of jets are merged.

\section{Results}
\label{sec:results}

First we study the effect of multi-jet production and azimuthal angle constraints in proton-proton collisions. We calculate the energy spectrum in CASTOR kinematics in the cases where 1 or two jets are merged, and an arbitrary number of other jets are produced but not merged with the two measured ones. These cross sections are calculated as shown in Eqs.~\eqref{eq:pp_1jet} and \eqref{eq:pp_2jet}.

\begin{figure}[tb]
        \centering
		\includegraphics[width=\columnwidth]{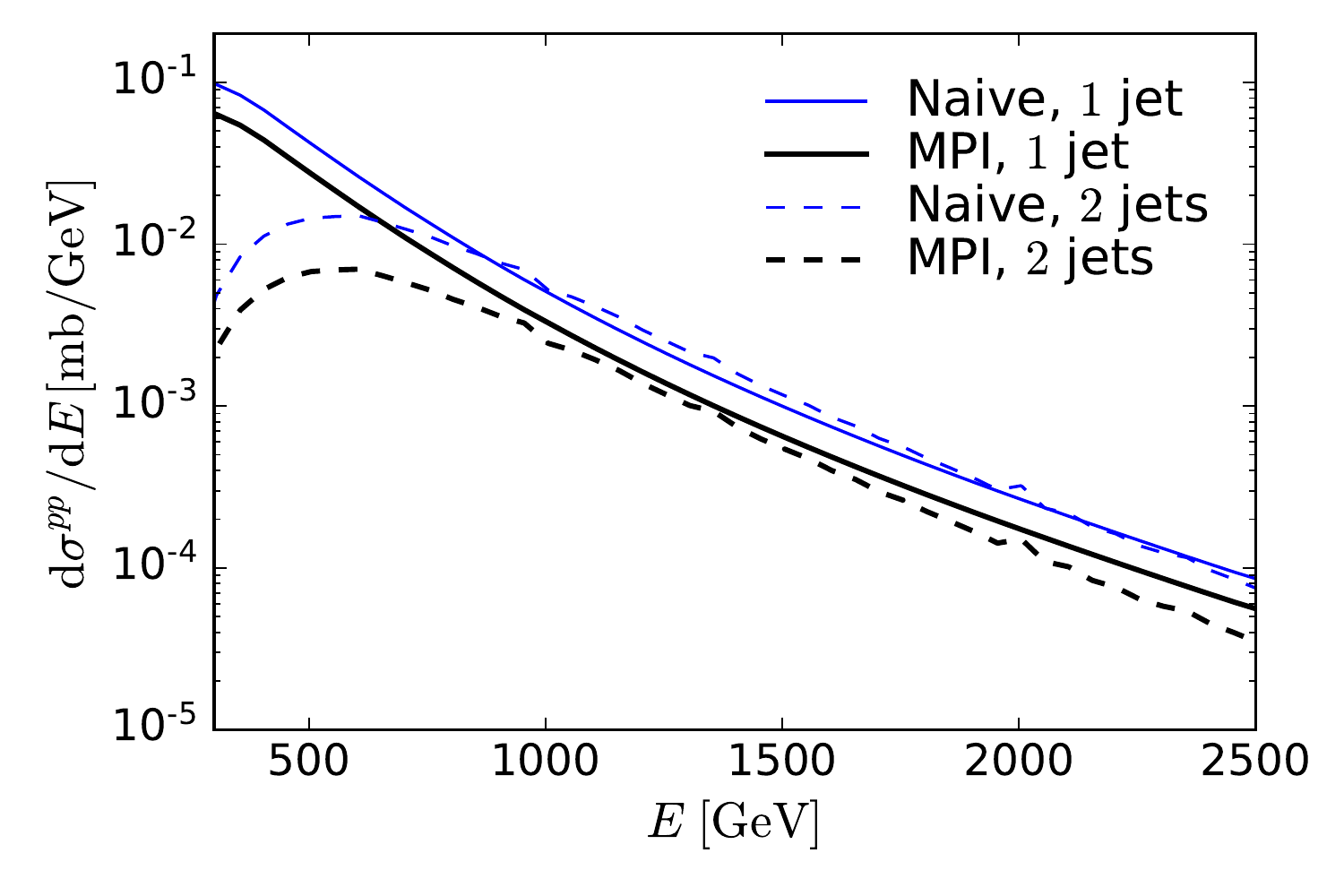} 
				\caption{ Jet energy spectra in proton-proton collisions at $\sqrt{s}=13$ TeV in the cases where one or two jets are merged. No rapidity shift is applied here. The thin lines refer to the ``naive'' approximation where no angular constraints are implemented on the jets that are not merged.
				}
		\label{fig:pp_mpi}
\end{figure}

For comparison, we show the single jet production spectrum in the ``naive approximation'', in which case  we do not put any constraints on the other jets and set $C_1(k)=1$. 
In case of two-jet production, this corresponds to the standard DPS result of Eq.~ \eqref{eq:dps_pp}, with the kinematical constraint $x_1+x_2<1$ included, but imposing no requirements on the other partonic processes that may take place simultaneously with the DPS process. Energy spectra in these cases are shown in Fig.~\ref{fig:pp_mpi} as thin blue lines, and should be compared to the thick black lines that correspond to our main result. 
The explicit treatment of multi-parton interactions is found to be necessary, as the constraints on the azimuthal distribution of jets reduce the single-jet production by $\sim 25\%$ and the two-merged-jet cross section by $\sim 40\%$. 
Note that this effect depends on the integrated cross section and thus on the low-$p_T$ cut. However, as discussed in Appendix \ref{app:ptcut}, the jet-production cross section depends only weakly on the infrared cutoff in the CASTOR kinematics. 

\begin{figure}[tb]
        \centering
		\includegraphics[width=\columnwidth]{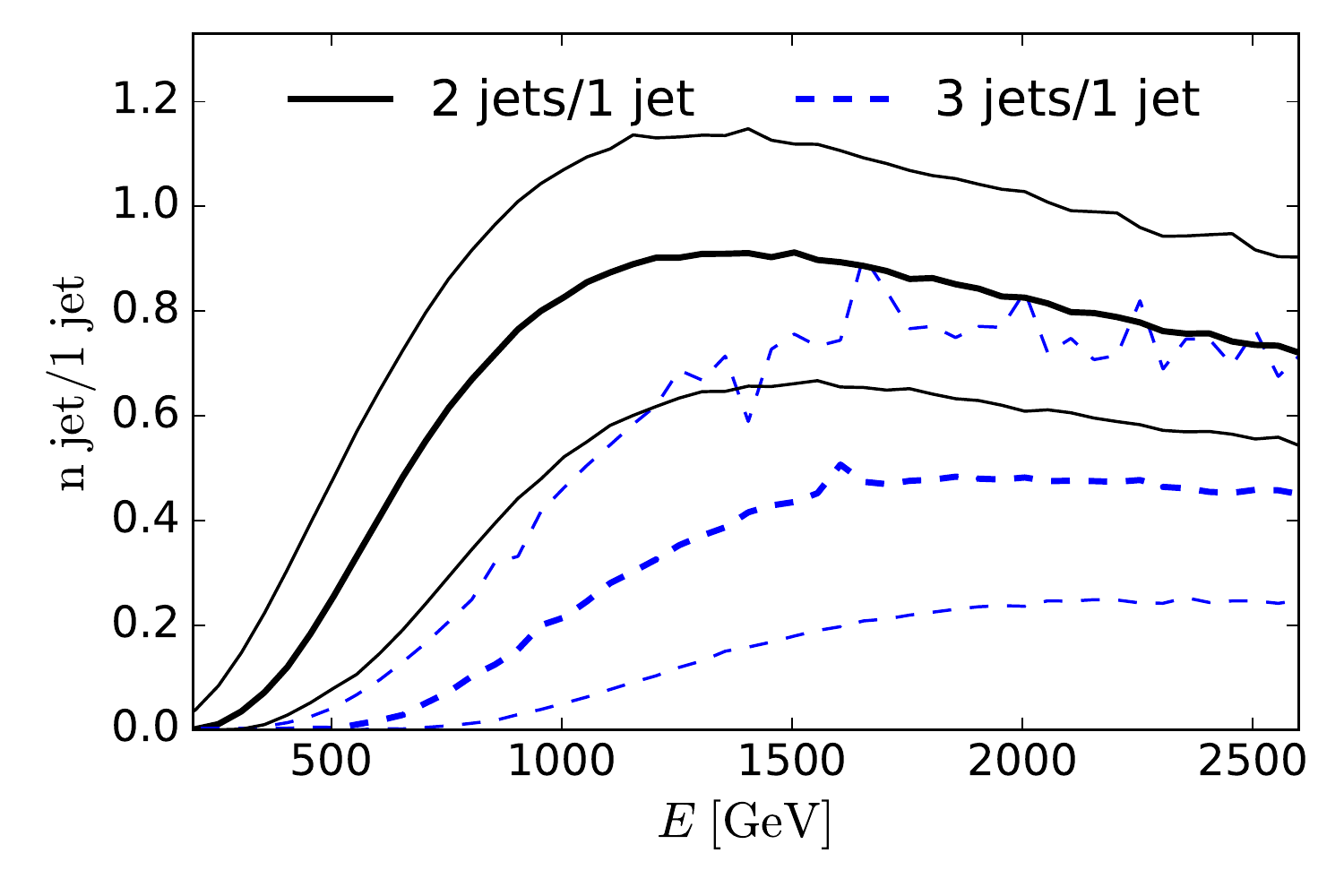} 
				\caption{Contribution to the jet production cross section in proton-nucleus collisions at $\sqrt{s}=5.02$ TeV from the cases where $n$ jets are merged into one. The results are normalized by the corresponding cross section in single jet production (with no merging) in proton-nucleus collisions. The thick lines are obtained with our standard choice for the infrared cutoff $p_{T,\text{min}}=1.0\gev$. Upper thin lines correspond to $p_{T,\text{min}}=0.5\gev$ and lower thin lines $p_{T,\text{min}}=1.5\gev$. }
		\label{fig:2_3_PS_ratio}
\end{figure}

Let us then study proton-nucleus collisions and  the contributions where two or more jets are merged into one in the CASTOR kinematics. To quantify this, we show in Fig.~\ref{fig:2_3_PS_ratio} the cross section $\der \sigma^\text{pA,n}/\der E$ in proton-nucleus collisions in case of $n$ jets merging, normalized by the single jet production cross section ($n=1$). 
The contribution from the two merged jets is found to be numerically significant, comparable with the single jet production cross section at $E\gtrsim 1$ TeV. On the other hand, contribution from the three merged jets is important only at highest energy bins $E\gtrsim 1.5$ TeV. 
When the infrared cutoff $p_{T,\text{min}}$ is lowered from our standard choice of $1.0\gev$ to $0.5\gev$ the multi-parton production processes become more important as the potential phase space becomes larger, and the multi- parton processes in general are more likely when the integrated cross section increases. Similarly, with $p_{T,\text{min}}=1.5\gev$ we find that multi-jet production is suppressed, and especially the three merged jet contribution becomes negligible. We will return to the dependence on the   minimum $p_T$ cut in Appendix \ref{app:ptcut}.

\begin{figure}[tb]
        \centering
		\includegraphics[width=\columnwidth]{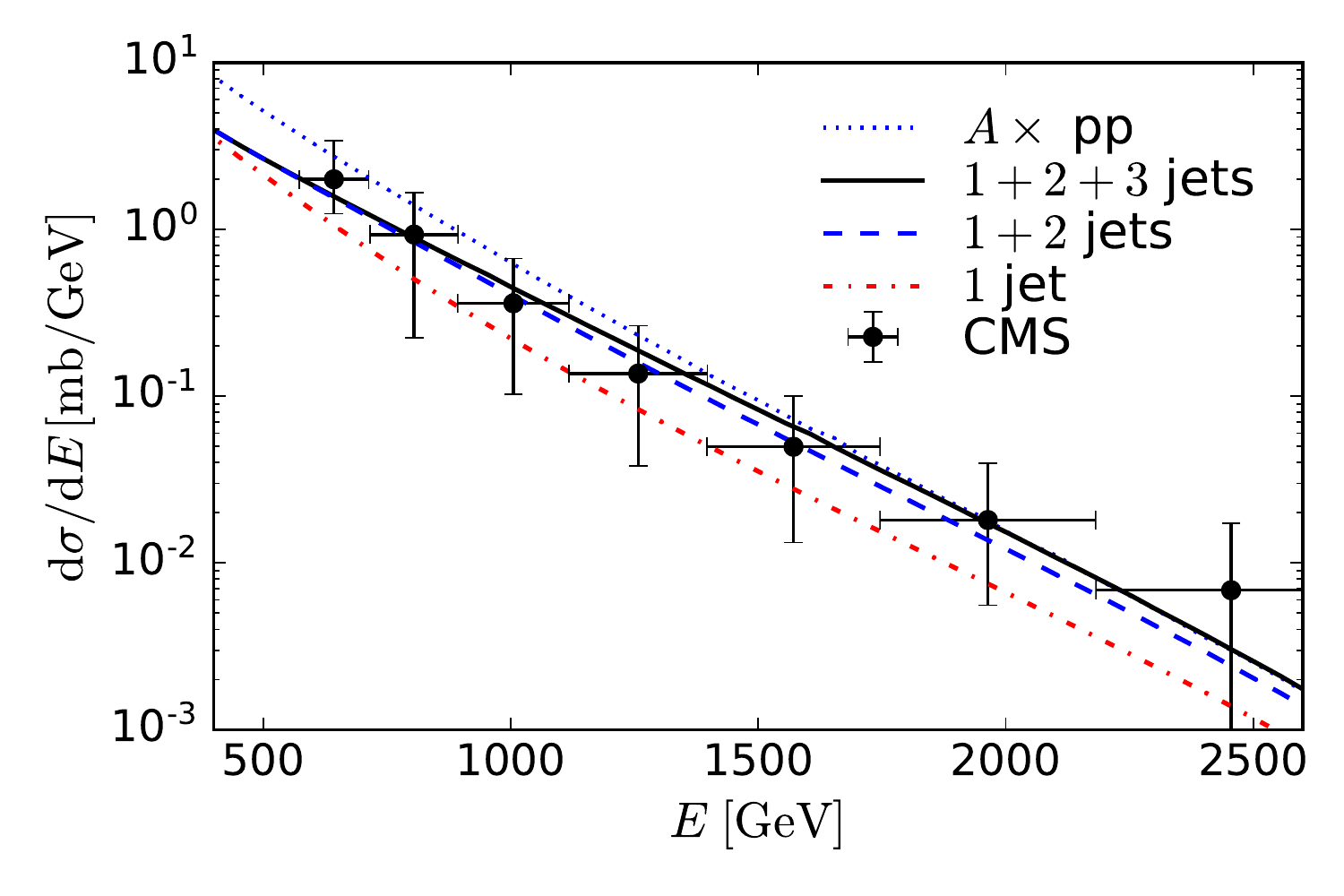} 
				\caption{Jet energy spectrum in proton-nucleus collisions compared with the CASTOR data~\cite{Sirunyan:2018ffo} at $\sqrt{s}=5.02\tev$ including one, two and three merged jet contributions. The scaled proton-proton result includes two- and three merged jets contributions.}
		\label{fig:pA_spectra_dps}
\end{figure}

The jet energy spectrum in proton-nucleus collisions is shown in Fig.~\ref{fig:pA_spectra_dps}. As already seen above, the contribution from the case where more than one jets are merged into one is significant. However, including the three jet merging contribution to the calculation is not a large correction anymore, and we can expect that contribution from the cases where more than $n=3$ jets are merged is negligible in comparison to the current experimental precision. Thus, we do not include these contributions which would be numerically more demanding. Overall, we find a good agreement with the  CMS data~\cite{Sirunyan:2018ffo} if the contribution from multi-jet production is included.

\begin{figure}[tb]
        \centering
		\includegraphics[width=\columnwidth]{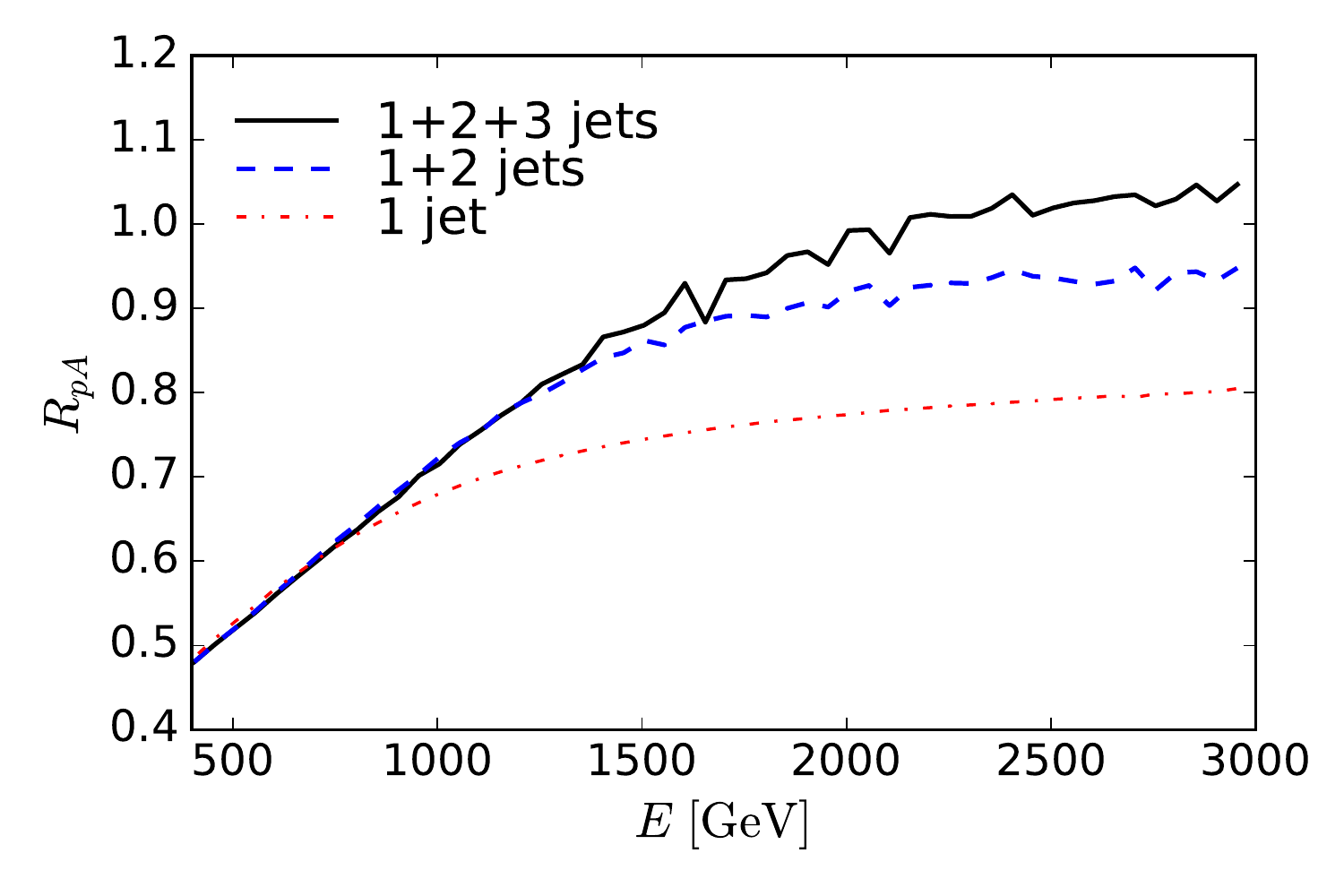} 
				\caption{Nuclear modification factor for the jet energy spectrum in CASTOR kinematics at $\sqrt{s}=5.02\tev$ including contributions from up to three merged jets. The proton-proton reference is computed with the same kinematical shift as applied in the proton-nucleus scattering. }
		\label{fig:rpa}
\end{figure}

To study the importance of the saturation effects, we compute the nuclear suppression factor 
\begin{equation}
R_{pA} = \frac{ \der \sigma^{pA} / \der E} {A \der \sigma^{pp}/\der E }
\end{equation}
Note that in $R_{pA}$ the proton-proton reference should be evaluated in the same kinematics, and thus we will also apply the boost $y_\text{shift}$ in this case, even though there is no proton-proton measurement done in this kinematics. The resulting nuclear suppression factor is shown in Fig.~\ref{fig:rpa}, and the results are  computed with and without multi-parton contribution. The nuclear suppression is found to be significant in the region where CASTOR data are available, reaching down to $R_{pA}\approx 0.5$ at smallest-energy bins. At $E \gtrsim 1$ TeV contributions from more than one merged jets becomes important and  $R_{pA}$ eventually becomes larger than unity, thanks to the likely multi-parton scatterings~\cite{Strikman:2001gz,Blok:2012jr,dEnterria:2012jam,Cazaroto:2016nmu,Helenius:2019uge}.

Fig.~\ref{fig:rpa} quantifies the expected magnitude of nuclear effects, but is not experimentally measurable as it is not expected that proton-proton collisions will be performed at the LHC with different energies for the two proton beams. However, cross-section ratios are still beneficial as many experimental and theoretical uncertainties cancel. Thus, we calculate the cross-section ratio
\begin{equation}
\label{eq:rshift}
	R_\text{shift} =  \frac{\der \sigma^{p+A\to i+X}/\der E (y_\text{shift}=0.465) }{  A \der \sigma^{p+p\to i+X}/\der E (y_\text{shift}=0) }. 
\end{equation}

\begin{figure}[tb]
        \centering
		\includegraphics[width=\columnwidth]{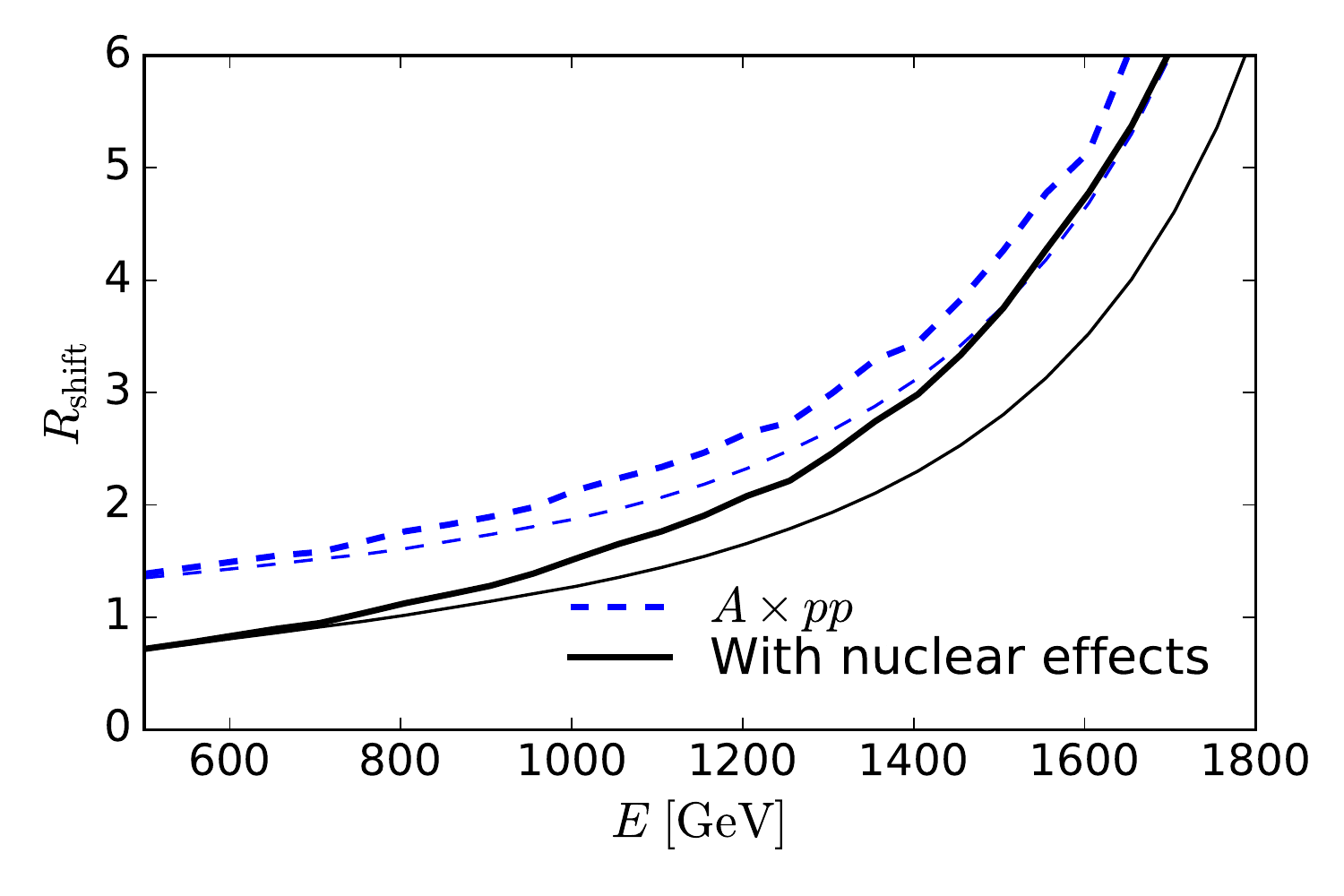} 
				\caption{Jet production cross section ratio $R_\text{shift}$ at $\sqrt{s}=5.02\tev$ defined in Eq.~\eqref{eq:rshift} in case of no rapdity shift in the proton-proton baseline with and without nuclear effects (see text for details). Thick lines include contribution from one,  two or three merged jets, and thin lines have no merged jets.  }
		\label{fig:rpa_shift}
\end{figure}

This ratio is studied in Fig.~\ref{fig:rpa_shift}  where we show $R_\text{shift}$ obtained with and without nuclear effects. No nuclear effects here means that we take the full cross section in proton-proton collisions (including contributions from merged jets), and scale by $A$. Note that in this case the multiparton interactions are not enhanced by the larger integrated cross section as is the case when we do full proton-nucleus calculation with nuclear effects. Consequently, two and three merged jet contributions have smaller effect in the case with no nuclear effects, as can be seen by comparing thin (one jet) and thick (up to three merget jets) curves in Fig.~\ref{fig:rpa_shift}. As already seen when studying $R_{pA}$ in Fig.~\ref{fig:rpa}, the nuclear effects suppress the ratio by a factor $\sim 2$ at small jet energies. This suppression is not significantly affected by the inclusion of merged-jet contributions. Similarly as in case of $R_{pA}$, the contribution from two or three merged is important at $E \gtrsim 1$ TeV. At large energies the cross section ratio diverges as as the reference spectrum dies much faster as a function of energy without the rapidity shift.  

\section{Conclusions}
We have calculated jet-energy spectra in proton-nucleus collisions at forward rapidities, in the kinematics covered by the CASTOR calorimeter at CMS. We demonstrate that the saturation effects that affect the nuclear structure at the probed Bjorken-$x$ values $x_A \sim  10^{-5} \dots 10^{-6}$ are large, resulting in a nuclear suppression by a factor $\sim 1/2$. The simple framework applied in this work, in which jet energy is approximated by the energy of the produced parton and a streamlined jet algorithm is applied, is compatible with the CMS results. The best agreement is obtained if we include contributions where the energies of two or three jets are summed when they are produced close to each other and their total energy is seen in the calorimeter. 

In this work we made two simplifying assumptions. First, we neglected higher- order corrections to the particle production cross section, as NLO calculations in the CGC framework are currently not developed to the level where phenomenological applications are straightforward. However, the BK evolution with running coupling is included, which resums large logarithmic corrections $\sim \as \ln 1/x$ to all orders.  Additionally, we neglected all fragmentation effects and assumed that the total parton energy is measured as a jet energy. One possibility to go beyond this simple assumption would be to include realistic fragmentation effects following e.g. Refs.~\cite{Deng:2014vda,Albacete:2016tjq} where the Lund string model based Monte Carlo methods were used to describe the fragmentation processes. A more realistic production mechanism for the jet formation would also motivate to use a more realistic jet finding algorithm to distinguish jets that are merged from the other jets whose energy is not included in the total energy of the observed jet. In the cross-section ratios like $R_{pA}$ or $R_\text{shift}$ model uncertainties should be significantly reduced, and it might be possible to also cancel many experimental uncertainties that are larger than the estimated saturation effects in case of inclusive jet spectra. Consequently, we argue that our main finding that the CASTOR measurements are sensitive to (and point towards) sizeable saturation effects is robust.

\section*{Acknowledgements}
We  thank K.~J.~Eskola, I. Helenius and P. Paakkinen for discussions, and T. Lappi for carefully reading  the manuscript. This work was supported by the Academy of Finland, projects 314764 (H.M) and 297058 and 308301 (H.P), and by the European Research Council, Grant ERC-2015-CoG-681707 (H.M). The content of this article does not reflect the official opinion of the European Union and responsibility for the information and views expressed therein lies entirely with the authors. Computing resources from CSC -- IT Center for Science in Espoo, Finland and from the Finnish Grid and Cloud Infrastructure (persistent identifier
   \texttt{urn:nbn:fi:research-infras-2016072533}) were used in this work.

\appendix

\vspace{5em}
\section{Dependence on the jet mass}
\label{appendix:mass}

Throughout this work we used $m=0.2\gev$ mass when computing the parton (jet) energy and mapping rapidity to pseudorapidity. In order to demonstrate that our results are not sensitive to this parameter, we calculate the energy spectrum $\der \sigma^\text{jet}/\der E$  in proton-lead collisions with  heavier mass $m=2.0\gev$, and compare the resulting spectra with our main result obtained with $m=0.2\gev$. 

\begin{figure}[H]
        \centering
		\includegraphics[width=\columnwidth]{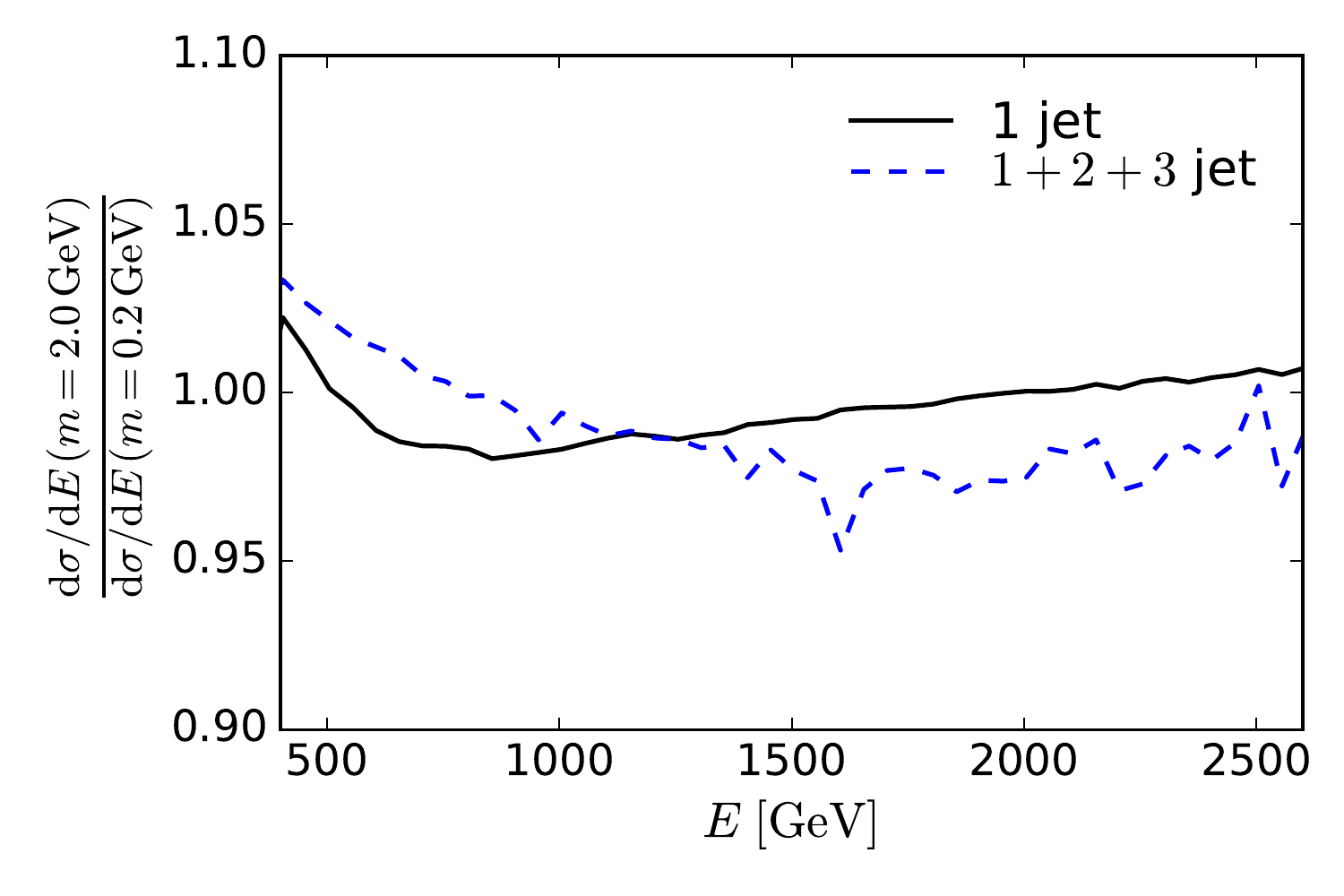} 
				\caption{Jet production cross section proton-lead collisions with  jet mass $m=2\gev$ relative to the same cross section computed with our standard choice of $m=0.2\gev$. Results with no merged jets (``1 jet'') and $1+2+3$ merged jets are shown. }
		\label{fig:mdep}
\end{figure}

The ratio of the cross sections obtained with different parton masses in proton-nucleus collisions is shown in Fig.~\ref{fig:mdep}, both in the case where we do not include the jet merging, and in the case where $1$, $2$ or $3$ jets are allowed to be merged. We find that the cross sections change by less than $5\%$ when the mass parameter is varied, which is much below both the model uncertainty and the experimental accuracy.

We have checked that the $R_{pA}$ and $R_\text{shift}$ are basically identical with all studied mass values.

\section{Dependence on the minimum $p_T$ cut}
\label{app:ptcut}
When integrating over the parton level kinematics in Eq.~\eqref{eq:jet_spectra},  we have to include a low-$p_T$ cut as the formalism is not applicable at non-perturbatively small momentum transfers. However, the jet-energy spectrum is an infrared safe quantity, and should depend weakly on this cut.

\begin{figure}[H]
        \centering
		\includegraphics[width=\columnwidth]{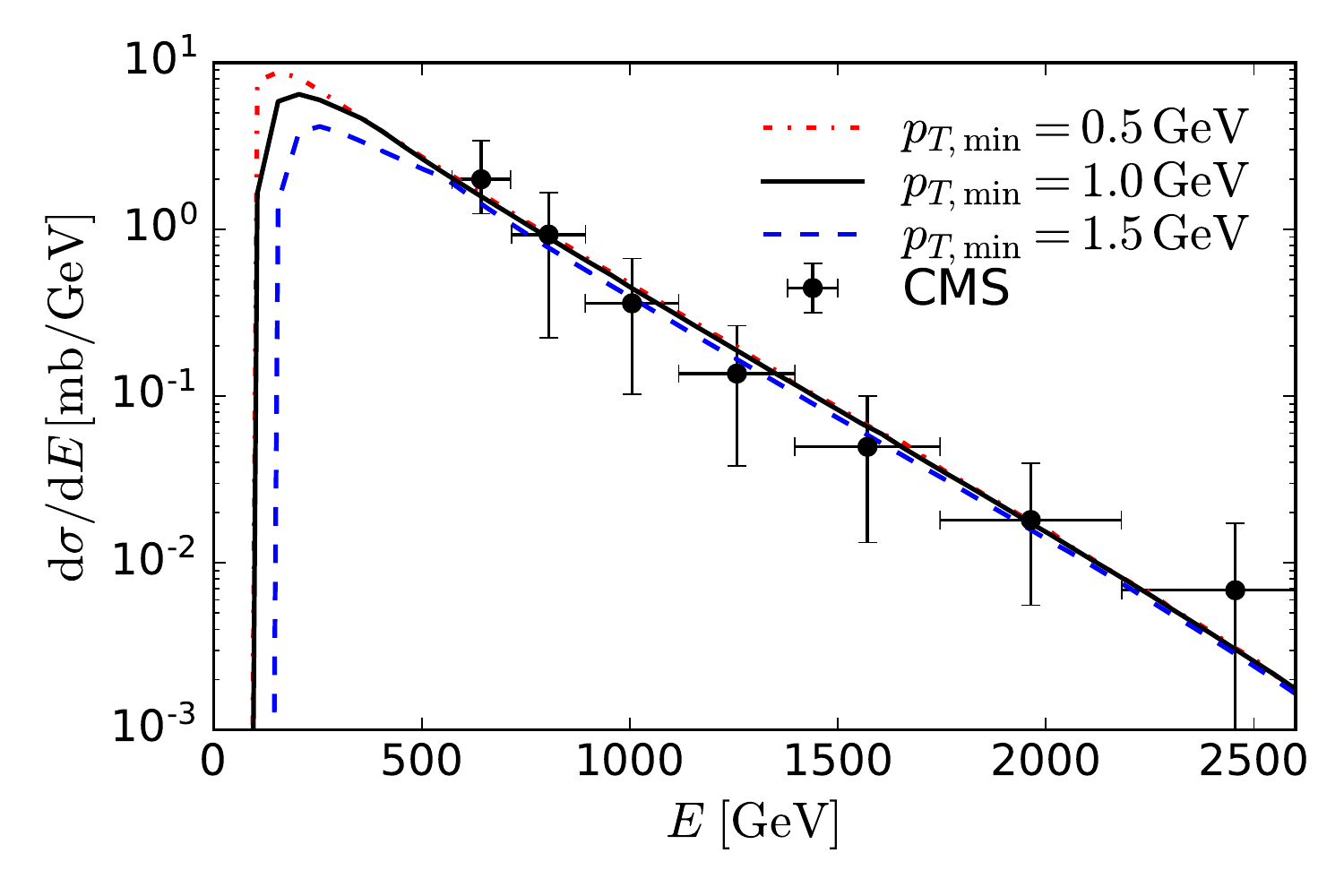} 
				\caption{Jet energy spectrum in proton-nucleus collisions with different lower $p_T$ cut. We include contributions from 1, 2 or 3 merged jets. }  
		\label{fig:ptcutdep}
\end{figure}

The dependence of the energy spectrum on different low-$p_T$ cuts is shown in Fig.~\ref{fig:ptcutdep}, where the jet energy spectrum in proton-lead collisions at $\sqrt{s}=5.02$ TeV is shown. As expected, varying $p_T$ cut around our standard choice of $p_{T,\mathrm{min}}=1.0\gev$ changes the spectrum only at lowest energies much below the CASTOR data.

\begin{figure}[H]
        \centering
		\includegraphics[width=\columnwidth]{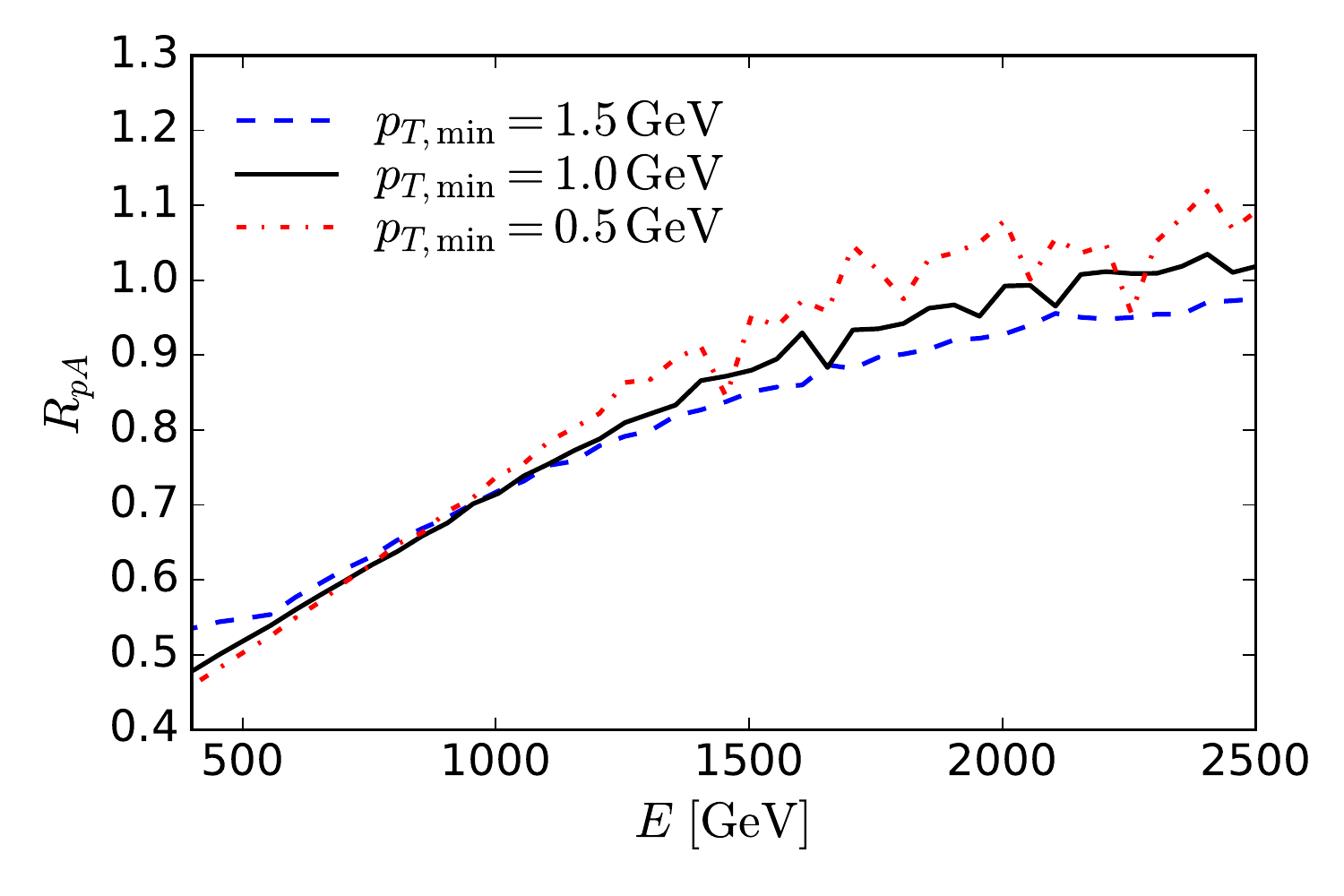} 
				\caption{Nuclear suppression factor calculated with two different infrared cutoffs. Contributions from 1, 2 and 3 merged jets are included. }  
		\label{fig:rpa_minpt}
\end{figure}

Next we demonstrate that the nuclear suppression factor, and consequently our estimate for the strength of the saturation effects in the CASTOR kinematics, neither depends strongly on the infrared cutoff. In Fig.~\ref{fig:rpa_minpt} we show the nuclear suppression factor $R_{pA}$ calculated with different minimum $p_T$ cutoffs. The contribution from 1, 2 or 3 merged jets is included. The resulting nuclear suppression factors are found to be comparable at all energies.

\section{Momentum conservation in the probe}
\label{appendix:kc}
Calculation of contributions where multiple jets are merged into one applies an effective descriptions for the double and triple parton distribution functions 
as written in Eq.~\eqref{eq:dpdf}. 
These functions are constrained such that they implement a longitudinal momentum conservation $\sum_{i=1}^n x_i<1$, where $n$ is the number of merged jets. At the forward rapidities in the CASTOR kinematics, the Bjorken-$x$ in the probe is large, and the energy conservation is expected to have a large effect. In this Appendix the importance of this kinematical constraint is demonstrated.

\begin{figure}[tb]
        \centering
		\includegraphics[width=\columnwidth]{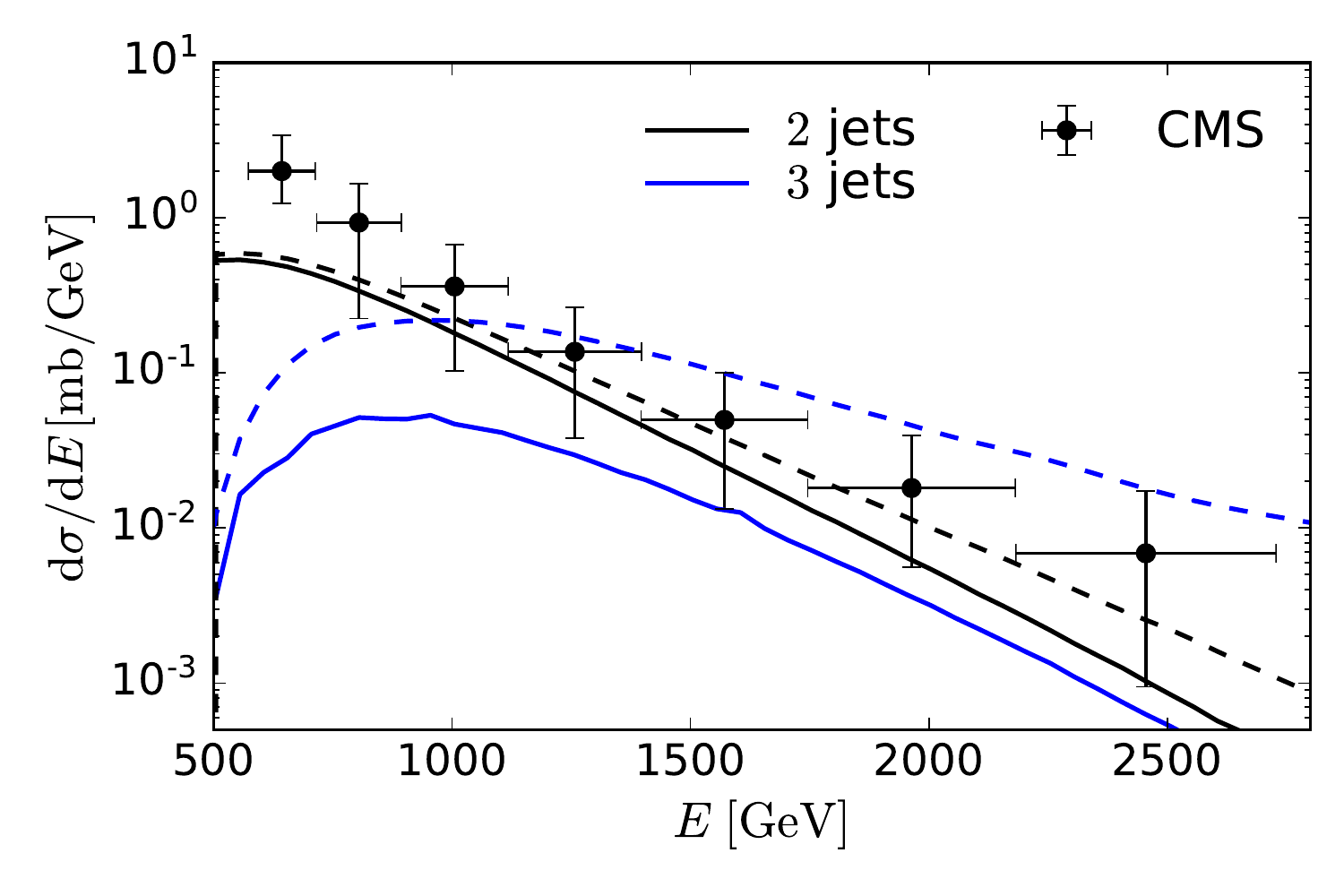} 
				\caption{Contribution on the jet energy spectra from two merged jets (black lines) and three merged jets (blue lines) in proton-nucleus collisions at $\sqrt{s}=5.02\tev$. Solid lines include the kinematical constraint, and dashed lines do not. }  
		\label{fig:kc}
\end{figure}

In the case where we have no merging, the kinematical constraint has no effect as the longitudinal momenta carried by the other $k$ jets that are not merged is neglected. In Fig.~\ref{fig:kc} the contribution to the jet energy spectra from two and three merged jets is shown, with and without the kinematical constraint (the dashed lines refer to the case where the kinematical constraint is not included). The cross section without a kinematical constraint is calculated by assuming that the multi parton distribution factorizes, e.g. for the two parton distribution we write 
 $  D_{ij}(x_1,x_2)=x_1f_i(x_1) x_2 f_j(x_2),$
without introducing the requrement $x_1+x_2<1$. Similarly the three parton distribution function is a product of three parton distribution functions. 

In case of two merged jets, the effect of the kinematical constraint is found to be moderate in the studied kinematics. On the other hand, when the number of merged jets increases to three, the kinematical constraint starts to have a clearly larger effect as expected. Below $E \lesssim 800 \gev$ the kinematical constraint has a small effect on the total jet spectra (including $n=1,2$ and $3$ merged jets), which suggest that in the region where we argue the saturation effects to be strong, our results are not sensitive to the details of how the poorly constrained multi parton distribution functions are implemented.

\bibliography{refs}
\bibliographystyle{JHEP-2modlong}

\end{document}